\documentclass[aps,pra,preprint,groupedaddress]{revtex4}

\usepackage{epsfig}
\usepackage{color,graphicx}
\usepackage{siunitx}
\begin{document}

\title{Ionization of Rydberg H atoms at band-gap metal surfaces via surface and image states}
\author{E. So}
\affiliation{Department of Chemistry, University of Oxford, Chemistry Research Laboratory, Oxford OX1 3TA, United Kingdom}
\author{J. A. Gibbard}
\affiliation{Department of Chemistry, University of Oxford, Chemistry Research Laboratory, Oxford OX1 3TA, United Kingdom}
\author{T. P. Softley}
\affiliation{Department of Chemistry, University of Oxford, Chemistry Research Laboratory, Oxford OX1 3TA, United Kingdom}

\date{\today}

\begin{abstract}
Wavepacket propagation calculations are reported for the interaction of a Rydberg hydrogen atom ($n=2-8)$ with Cu(111) and Cu(100) surfaces (represented by a Chulkov potential), in comparison with a Jellium surface.  Both copper surfaces have a projected band gap at the surface in the energy range degenerate with some or all of the Rydberg energies.  The charge transfer of the Rydberg electron to the surface is found to be
enhanced for $n$ values at which there is a near-degeneracy between the Rydberg energy level and an image state or a surface state of the surface.  The enhancement is facilitated by the strong overlap of the surface image-state orbital lying outside the surface and the orbital of the incoming Rydberg atom. These calculations point to the possibility of using Rydberg-surface collisions as a probe of surface electronic structure.

\end{abstract}


\maketitle

\section{Introduction}
A Rydberg atom has one electron in a very weakly bound orbital, with a mean orbital radius that scales with $n^2$, where $n$ is the principal quantum number. When such an atom approaches a solid surface, the state is strongly perturbed at long range by fields originating from the surface, due to the van der Waals interaction between the atom and its own image dipole in the surface, and also to localized charges at the surface \cite{Sashi1,Sashi2,Neufeld:2010p1915}. Ultimately the perturbation may lead to charge transfer of the Rydberg electron to the surface -- the surface ionization of the Rydberg atom -- which typically occurs at a distance of around 4$n^2$ in atomic units (e.g., for $n=20$ the surface ionization distance is around 80 nm). In practice the ionization does not occur at a specific distance but with an increasing probability as the atom approaches the surface (or equivalently as a varying ionization rate as a function of distance from the surface).

An important question to ask is how much the ionization probability  associated with the charge transfer process depends on the nature of the surface. The target surface can be physically structured (e.g.\ a rough surface, a stepped or vicinal surface, a surface with adsorbates) or electronically structured (e.g.\ band-gap structure, or thin films or adlayers with embedded quantized states).  It may also have localized charges as discussed above.  For the charge transfer from a Rydberg atom to a conducting metal surface, the Rydberg energy is generally degenerate with the conduction band of the metal.  Hence the charge transfer is resonant, and it would not be expected that the band structure of the metal would have a significant effect on this process. So for example gold and aluminum would be expected to show broadly similar behavior.

However there are other situations where the surface does not have a true continuum of states at the appropriate energy for charge transfer. For example,  a thin metal film behaves as a quantized 1D well in the dimension perpendicular to the surface, and as a free-electron system (unquantized) in the dimensions parallel to the surface (thus forming a 2D band), as opposed to a Jellium surface which is effectively unquantized in all dimensions. Usman \textit{et al.} studied such quantization effects  on the charge transfer dynamics of ground state H$^{-}$ ions (from the negative ion to the metal surface)  \cite{Usman:2001p284}. It was found that when the electronic energy of the H$^-$ is close to that of a 1D quantized well state, the rate of charge transfer is enhanced by the resonance, and the resultant ionization probability as a function of distance was found to exhibit multiple peaks due to the many quantized thin film states that the atomic energy level crosses when the H$^-$ approaches the thin film \cite{Usman:2001p284}.

The observation of a resonance effect close to the 1D quantized level can be explained by the preferential direction of charge transfer, which is along the  $z$-axis on which the saddle point in the effective potential occurs, and so the charge transfer is most efficient when there is minimum transfer of momentum in the direction parallel to the surface, i.e., when the Rydberg energy is close to the 1D quantized state. An alternative perspective is that states with momentum parallel to the surface have high angular momentum with regard to the position of the approaching atom, and therefore electron transfer into these states is hindered by conservation of angular momentum.
A similar quantized-state resonance effect has been studied by Gauyacq and coworkers for  the charge transfer of H$^{-}$ ions at a Cu(111) metal surface, but in the context of surface and image states embedded in the projected band-gap \cite{Borisov:1999p1720,PRL1998,PRL84}, rather than quantized 1D well states for the case of thin films. 

In this article the effects of electronically structured surfaces are investigated for the charge transfer of the Rydberg hydrogen atom, with principal quantum number $n=2-8$, at the projected band gap surfaces Cu(100) and Cu(111). The advantage of studying a Rydberg system compared to ground state H$^-$ is that the resonance effects in charge transfer can be explored over a wide energy range by varying the principal quantum number of the Rydberg atom. Additionally high-energy image and surface states that lie in the energy range of experimentally accessible Rydberg hydrogen atoms can be probed. Such states are not easily probed by other methods. As is shown here, resonances occur at several values of $n$ and at different values for the two surfaces.  Wavepacket propagation calculations are presented that illustrate these resonant effects.   

\section{Theory}
\subsection{Surface states and image states}
Surface states and image states are closely related, having energies within an energy gap of the band structure associated with the direction of the surface normal ($\vec{z}$).  Their electronic wavefunctions tend to zero towards the bulk metal as well as towards the vacuum ($\Psi \rightarrow 0, z_e\rightarrow \pm \infty$, where $z_e$ is the electron position), and they are therefore mainly confined to the metal-vacuum interface. 
Intrinsic surface states arise from the cleavage of the bulk metal (translational symmetry perpendicular to the surface is lost), and are localized mainly in the surface atomic layer \cite{Tamm:1932p1829,Shockley:1939p1830}, with a wavefunction that decays exponentially towards the vacuum. Image states arise from the Coulomb-like attractive image potential for an electron outside the metal surface ($V(z_e)=-\frac{1}{4}(z_e-z_{\rm im})^{-1}$  where $z_{\rm im}$ is the image plane position) and from the surface barrier created by a gap of available bulk electronic states in the metal.  

Image states are localized mainly in the vacuum region of the interface (or more precisely,  where the electron position  is outside the image plane position  \cite{Echenique:1978p1825}.) The energies of the image states form a hydrogen-like Rydberg series (setting the usual nuclear charge term $Z$ as $1/4$):
\begin{equation}
E_{n_{\rm img}}^{\rm IS}=-\frac{1}{16} \cdot \frac{1}{2(n_{\rm img}+a)^2}~,
\label{eq:IS_en}
\end{equation}
where $n_{\rm img}$ is the image-state index and $a$ is the quantum defect parameter for a given surface (for the Cu(111) and Cu(100) surfaces studied below, $a$ is approximately 0.02 and 0.24 respectively \cite{Chulkov:1999p1710,Klamroth:2001p1713}). The corresponding 1D wavefunctions have the form \cite{Klamroth:2001p1713}
\begin{equation}
\Psi^{\rm IS}_{n_{\rm img}} (z_e) = z_e R_{n_{\rm img},l=0}(z_e /4),
\label{eq:IS_wf}
\end{equation}
where $R_{n_{\rm img},l=0}(z_{e}/4)$ is the normalized hydrogenic ($s$-wave) radial wavefunction. Thus, image states can extend far into the vacuum ($<z_{e}>^{\rm IS}_{n_{\rm img}} = 6(n_{\rm img}+a)^2$a$_0$), and in the context of the Rydberg-surface interaction there can be significant overlap with the Rydberg electronic wavefunction at a long distance from the surface. Thus, the resonance effects in charge transfer via image states are expected to be particularly significant, especially when compared with the surface states which reside primarily inside the surface.
For conducting metals such as gold, the image states are degenerate with the conduction band and the strong coupling of the image states to the degenerate continuum means that the energies are extremely broad and the resonant states short lived. Thus it is only for band-gap semiconductors (such as the Cu(111) and Cu(100) surfaces considered here) where such states are sufficiently narrow in energy and long lived to observe resonant charge transfer.
\begin{figure}
\includegraphics{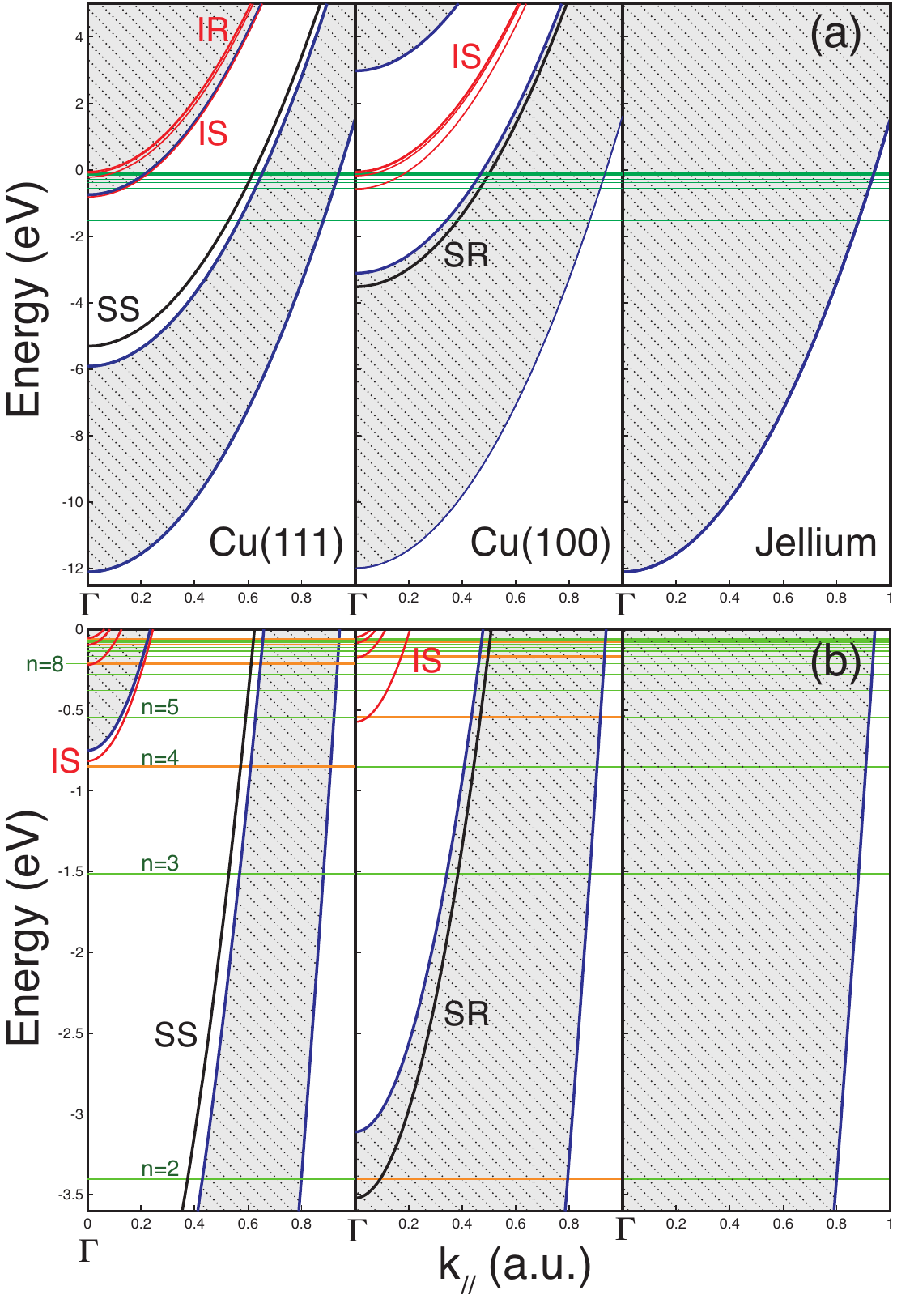}
\caption{(a) Energy of the electronic states in the model Cu(111), Cu(100) and free electron Jellium surface, as a function of electron momentum parallel to the surface, $k_{\parallel}$. The 3D bulk states are hatched and shaded in grey, the surface state (SS) and surface resonance (SR) are plotted as full black lines, image states (IS) and image resonance (IR) are plotted as full red lines, and the H atom Rydberg energies (1/2n$^2$ a.u.) are shown as green horizontal lines. (b) is the same as (a) but expanded in the energy range of the Rydberg states of an external H atom. Rydberg levels close to a surface- or image-state or resonance are colored orange. The gamma point $\bar{\Gamma}$ at $k_{\parallel}=0$ is also marked. The diagrams are based on those from Ref.\  \cite{Chulkov:2006p135}.}
\label{fig:dispersion}
\end{figure}

For an uncorrugated surface, the electron moves quasi-freely parallel to the surface ($\vec{\rho}$), such that each surface state and image state forms a 2D continuum of states with energy
\begin{equation}
E^{\rm IS,SS}_{n_{\rm img}}(k_{\parallel})= E^{\rm IS,SS}_{n_{\rm img}} + \frac{k_{\parallel}^2}{2\mu}~,
\label{eq:2D_band_en}
\end{equation}
where the second term is the dispersion energy, with electron momentum parallel to the surface $k_{\parallel}$, and an electron of effective reduced mass $\mu \simeq$ 1 (in atomic units) \cite{BORISOV:2002p1827}. Figure~\ref{fig:dispersion} shows the projected bulk band structure and the calculated energies of the surface and image states of the Cu(111) and Cu(100) surfaces \cite{Chulkov:2006p135}. Note that the full 3D bulk states are shaded in grey, and the projected band-gap is unshaded. Surface and image states that lie outside the energy gap, and are degenerate with the 3D bulk continuum, are commonly referred to as surface and image resonances (rather than `states') and are broadened by the coupling to the continuum.

\subsection{Wavepacket propagation calculations}

The details of the implementation of the wavepacket propagation approach for the work presented here are given in Ref.~\cite{So:2009p199}. 
In brief, the initial electronic wavefunction at atom-surface separations of $D_0\sim6n^2$ is found through matrix diagonalization using a Lagrange-Laguerre Discrete Variable Representation (DVR) for the radial coordinate \cite{Baye:2002p527}, and a Legendre DVR for the angular coordinate, $\theta$ (the metal surface is assumed to be flat and isotropic in dimensions parallel to the surface plane, thus the system is cylindrically symmetric and rendered two-dimensional). The radial part of the initial electronic wavefunction is then projected onto a Coulomb Wave DVR \cite{Dunseath:2002p1357} where the time propagation takes place. 

At each time step, the electronic wavefunction at a later time (or distance from the surface) is computed through the action of the time evolution operator, $e^{-i\hat{H}\Delta t}$:
\begin{equation}
\Psi_{\rm el}(\mathbf{{r}};t+\Delta t)= e^{-i\hat{H}_{\rm el}(\mathbf{{r}};t)\Delta t} \Psi_{\rm el}(\mathbf{{r}};t),
\label{eq:1}
\end{equation}
where  $\hat{H}_{\rm el}(\mathbf{{r}};t)$ is the electronic Hamiltonian which contains the kinetic energy operator, the centrifugal potential, and potential terms due the Coulombic interaction with the ion core, the applied electric field and the interaction of the electron with the surface. The time evolution propagator is approximated by a symmetric split operator \cite{FEIT:1983p1411} and the details are given in Ref.~\cite{So:2009p199}. 
In most of these calculations, the surface is moved towards the H atom at constant velocity 
and, unless otherwise stated, the velocity is taken to be $v_{\perp} =  3\times10^{-4}~\mathrm{a.u.}~\equiv656 \si{\meter \per \second}$ which is comparable to the mean velocity component perpendicular to the surface in experiments in our laboratory \cite{So:PRL,Cuexpt}.

In order to impose the correct outgoing boundary condition and to remove the spurious reflections of the wavefunction from the edge of the numerical grid, a complex absorbing boundary is included close to the grid edge \cite{GonzalezLezana:2004p1835}. To study the extent of the Rydberg-surface charge transfer, the total electron density on the numerical grid bound by the absorbing boundary is monitored as a function of atom-surface separation, and its derivative with respect to atom-surface separation corresponds to the ionization probability at the given distance.

The advantage of using the CWDVR is its suitability for treating the Coulomb potential (in particular the singularity at the origin) and for describing the  wavefunction in the metal, with a dense distribution of grid points near the origin and becoming regularly spaced at long range. This, together with favorable scaling of the wavepacket time-propagation (which only involves matrix multiplications, Eq.~\ref{eq:1}), allows calculations over a  range of principal quantum numbers to become feasible, albeit still very computationally demanding. 

\subsection{One-electron pseudo-potential}
\begin{figure}
\includegraphics[scale=0.45]{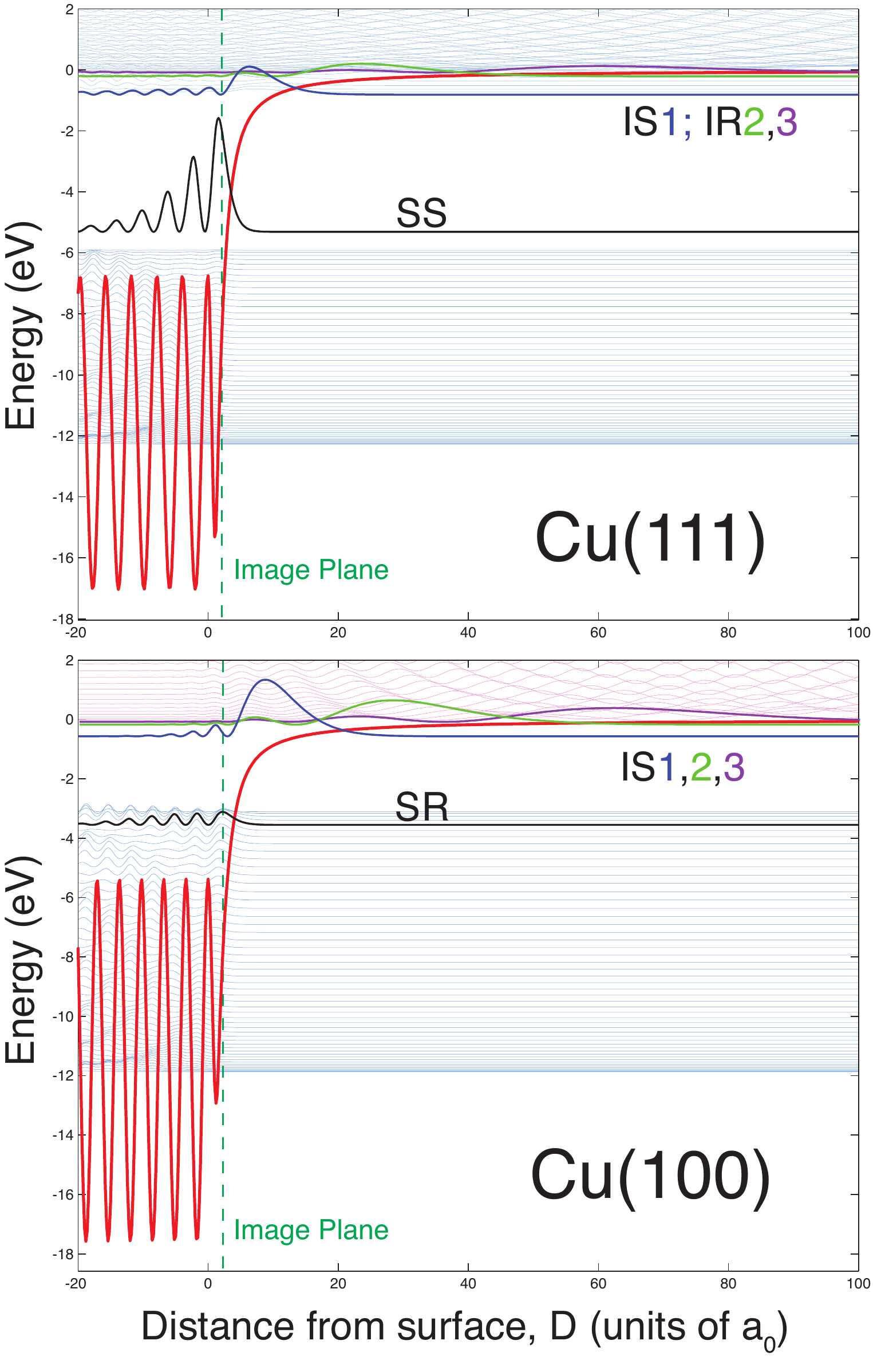}
\caption{Energies and wavefunctions of the bulk, surface and image states for the Cu(111) and Cu(100) surface, calculated from the diagonalization of the 1D pseudopotential given by Eq.~\ref{eq:chulkov} with a Sinc DVR basis \cite{LILL:1982p1364}. The energies and wavefunctions are the eigenvalues and eigenvectors respectively obtained from the matrix diagonalization. SS, SR, IS, IR labels the surface state, surface resonance, image state and image resonance respectively.}
\label{fig:Cu100_111}
\end{figure}

Following the work of references \cite{Borisov:1999p1720,Borisov:2001p1240,BORISOV:2002p1827,Klamroth:2001p1713}, the one-electron pseudo-potential of Chulkov \emph{et al.} \cite{Chulkov:1999p1710} is used to model the Cu(111) and Cu(100) surfaces in the direction normal to the surface ($\vec{z}$), while the electron is allowed to move freely in the direction parallel to the surface ($\vec{\rho}$). The potential has the analytical form:
\begin{equation}
\label{eq:chulkov}
\begin{array}{rlll}
V_{ee}(z_e)& = &V_1(z_e)+V_2(z_e)+V_3(z_e)+V_4(z_e)&   \\
V_1 (z_e)  &=  &A_{10} +A_{1}\cos \left( \frac{2\pi}{a_s}z_e \right) &  z_e<0  \\
V_2 (z_e)  &=  &-A_{20} +A_{2}\cos [ \beta(z_e) ]  &0<z_e<z_1  \\
V_3 (z_e)  &=  &A_{3} \exp[-\alpha(z_e-z_1)]  &z_1<z_e<z_{\rm im} \\
V_4 (z_e)  &=  &\frac{\exp[-\lambda(z_e-z_{\rm im})]-1}{4(z_e-z_{\rm im})} & z_{\rm im}<z_e ~, \\
\end{array}
\end{equation}
where $a_s$ is the bulk interlayer spacing, $A_{10},A_{1},A_{2}$ and $\beta$ are the independent model potential parameters, and the values for the Cu(111) and Cu(100) surfaces are given in Table~\ref{tab:chulkov}. 
\begin{table}
	\centering
		\begin{tabular*}{1\textwidth}{@{\extracolsep{\fill}}*{6}{c}}
		\hline  \hline
   	    & $a_s$ (a.u.) & $A_{10}$ (eV) & $A_{1}$ (eV) & $A_{2}$ (eV) & $\beta$ (a$_0^{-1}$) \\
   	   \hline
  		 Cu(111)  & 3.94 & -11.895 & 5.14 & 4.3279 & 2.9416 \\
             \hline
  		 Cu(100)  & 3.415 & -11.480 & 6.10 & 3.7820 & 2.5390 \\
  	 \hline \hline
		\end{tabular*}
\caption{Values of the parameters used in the one-electron pseudo-potential given by Eq.~\ref{eq:chulkov} for the Cu(111) and Cu(100) surface \cite{Chulkov:1999p1710}.}
\label{tab:chulkov}
\end{table}
The remaining parameters in Eq.~\ref{eq:chulkov} are determined by requiring the potential and its first derivative to be continuous:
\begin{equation}
\begin{array}{rlllrlll}
A_{20} &=& A_{2}-A_{10}-A_{1}
& ~~& A_{3} &=& -A_{20}-\frac{A_2}{\sqrt{2}}  \\
z_1 &=& \frac{5\pi}{4\beta} 
& ~~&\alpha &=& \frac{A_2 \beta }{A_3} \sin(z_1 \beta )  \\
\lambda &=& 2 \alpha  
& ~~&z_{\rm im} &=& z_1 - \frac{1}{\alpha} \ln \left( {\frac{-\alpha}{2A_3}} \right).
\end{array}
\end{equation}
The bulk potential and the width and position of the energy gap is described by $V_1$ in Eq.~\ref{eq:chulkov}, while the metal-vacuum interpolation potential and the energies of the surface and image states are described by $V_2$ and $V_3$. The $V_4$ term describes the long range image-potential. 
The form of the Cu(111) and Cu(100) pseudo-potential, and the energies and wavefunctions of the surface state and image states calculated from diagonalization of the 1D potential are shown in Fig.~\ref{fig:Cu100_111}. 
For comparison the calculations for these surfaces are benchmarked against a Jellium surface \cite{Cu_param} and the parameters used are those for aluminum.

\section{Results}
\subsection {Resonant charge transfer to surface states and image states}

The computation of the wavepacket propagation for the electronically structured Cu(111) and Cu(100) surfaces is much more demanding than the Jellium surface. This is due to the periodic nature of the pseudo-potential inside the metal, which requires a large number of radial grid points, and the large components of the electron momentum parallel to the surface in the resonant charge transfer (see below), which requires a large angular momentum basis. 
Typically for the calculations carried out below, the numerical grid consists of  $\sim1000 \times 200$ (radial $\times$ angular) grid points. 
Even so, for the charge transfers that are restricted by the band-gap and unaided by the resonance effects of surface or image states (see below), it is found that high frequency oscillations are superimposed on the  ionization probability profiles, and an increase of radial and angular points would alter the \emph{form} of the profiles. 
However, it was found that for the results presented below, the calculations are sufficiently converged that the average ionization distance, and the range of atom-surface separations spanned by the ionization probability profiles, do not change significantly with a larger radial or angular basis, and so are acceptable for the  comparison carried out below. 
Furthermore, it is shown below that resonant charge transfer to band-gap embedded surface states and image states can be observed directly from visual snapshots of the wavepacket propagation, confirming that the different onset of ionization distances is not an artefact of the unconverged calculations.

\begin{figure}
\includegraphics[scale=0.45]{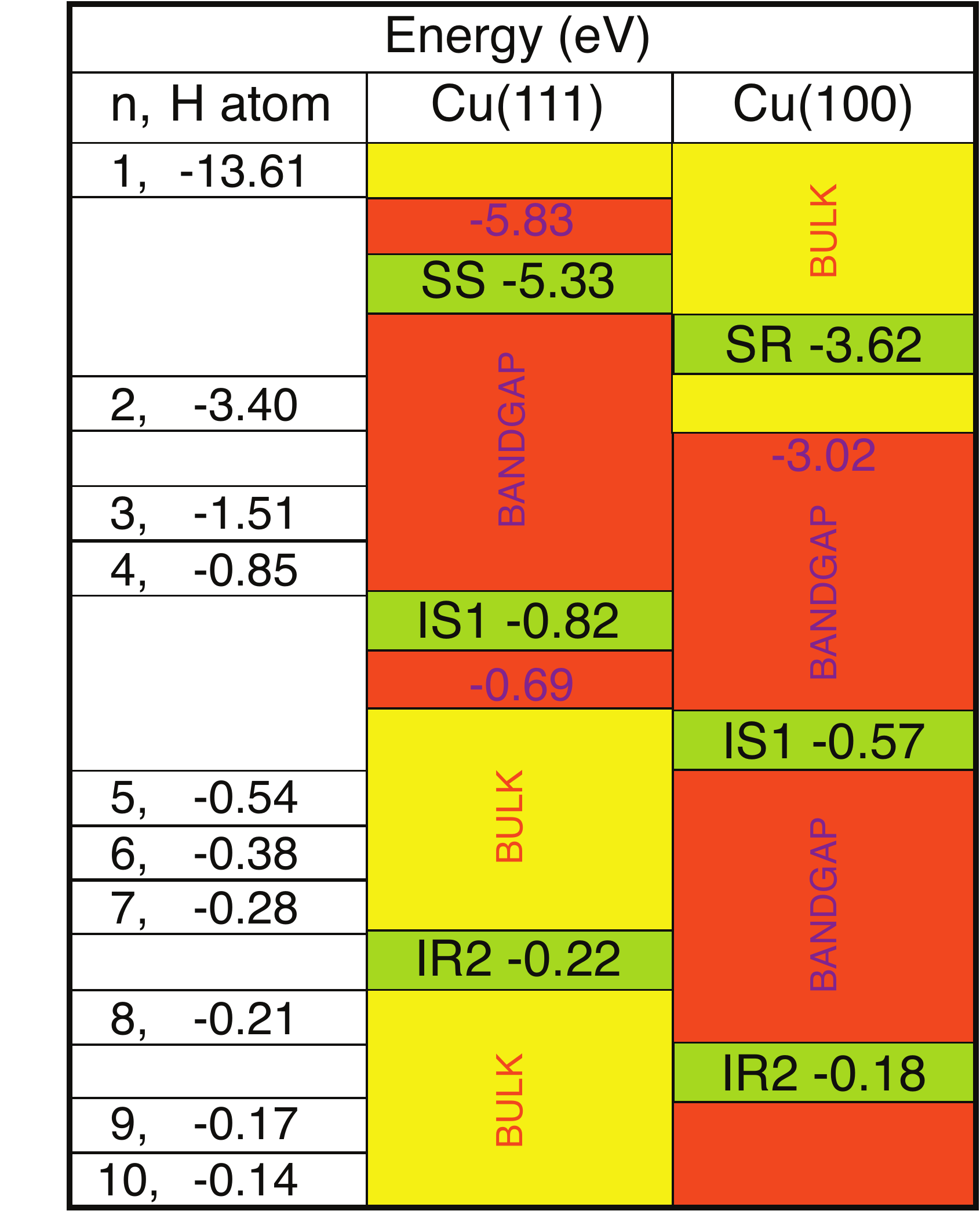}
\caption{Table of energies of the band gap, surface and image states or resonances for the Cu(111) and Cu(100) surfaces at the gamma point $\bar{\Gamma}$ ($k_{\parallel}=0$ ), compared to those of the Rydberg states of the hydrogen atom.  SS, SR, IS, or IR labels the surface state, surface resonance, image state and image resonance respectively.}
\label{fig:Cu_energytable}
\end{figure}

The calculations presented here are for the H atom for which the levels of a given $n$ have a degeneracy (arising from $l$ and $m_l$) of $n^2$.  
Under the surface potential (assumed cylindrically symmetric) the $l$-degeneracy of the Rydberg $n$-manifold is broken, and the Rydberg manifold splits and forms states which are polarized with respect to the surface normal \cite{Nordlander:1990p1802} akin to the Stark effect. If a field is present perpendicular to the surface (as typically occurs in experimental measurements \cite{So:PRL}, then the energy levels are also split at long range from the surface by the normal linear Stark effect, and as the atom 
approaches the surface  from long range these states correlate adiabatically with the surface-polarized states.
To focus on the resonance effects of the band-gap embedded surface states and image states, the calculations are performed here on the most surface-oriented state  (the most red-shifted state) of the $n$-manifold, which  has the largest overlap with the surface.  All the calculations reported here used an incoming velocity of $656~ \si{\meter\per\second}$ for the Rydberg atom along the surface normal.

\begin{figure}
\includegraphics[scale=0.45]{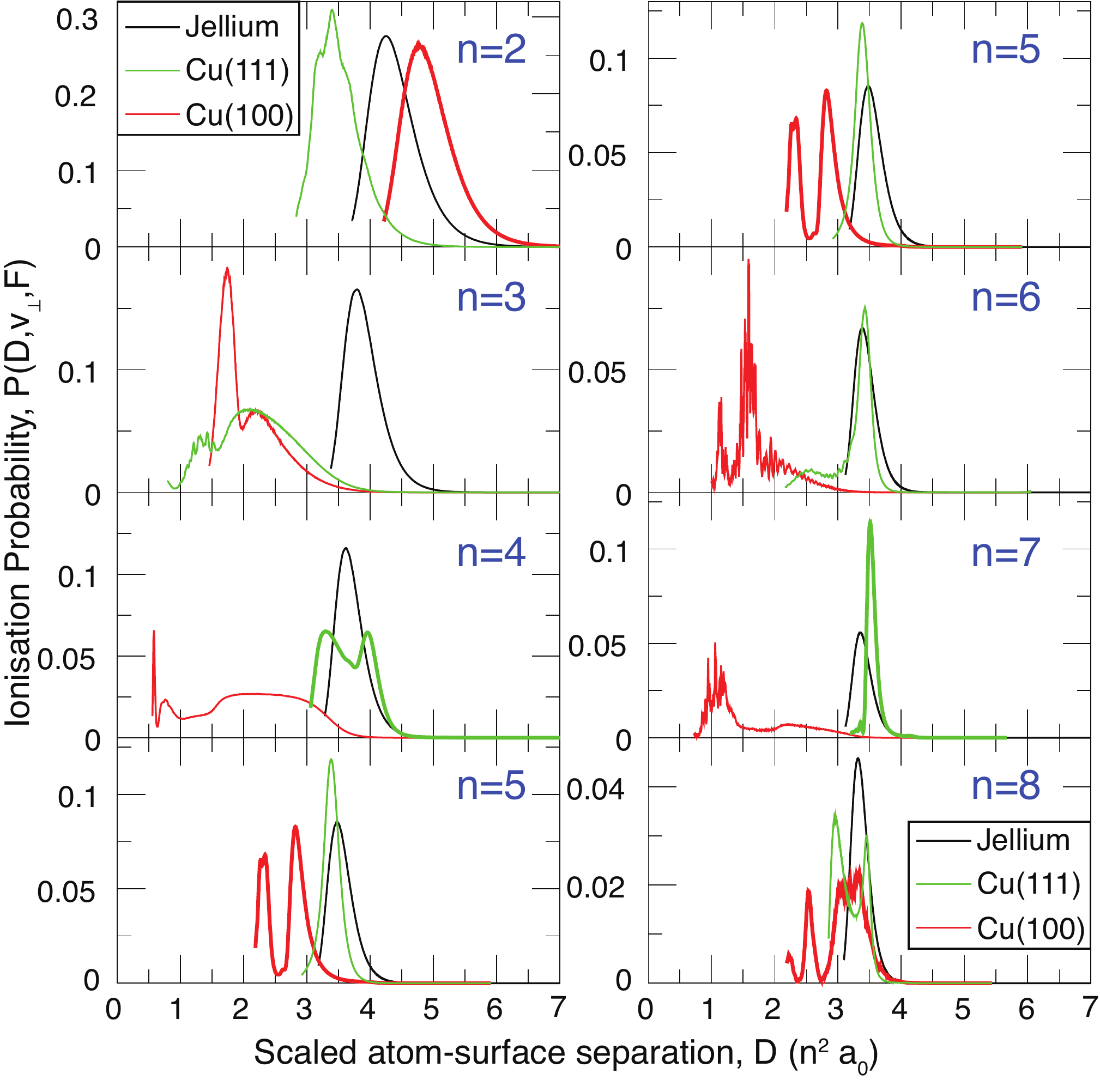}
\caption{Ionization probabilities as a function of \emph{scaled} atom-surface separation for the $n=2-8$ most surface-oriented Rydberg states calculated from the wavepacket propagation approach for a Jellium, Cu(111) and Cu(100) model potential. Note that the $n=5$ ionization probabilities are plotted twice to aid the comparison of the scaled ionization distances for each column. Rydberg states that are close to resonance with the surface or image states of Cu(111) and Cu(100) surfaces are shown in bold.}
\label{fig:Cun2-8}
\end{figure}

The energies of the unperturbed H atom states are shown in Fig.~\ref{fig:dispersion} as horizontal lines to compare with the energies of Jellium, Cu(111) and Cu(100) surface electronic states. The values of the energies are given in Figure~\ref{fig:Cu_energytable} to provide a more quantitative comparison and these suggest that there might be two or three resonances between the image or surface states and the H atom levels in the range $n=2-8$.
Figure~\ref{fig:Cun2-8} shows a comparison of the $m_{l}=0$, $n=2-8$ wavepacket propagation ionization probabilities calculated for a Jellium, Cu(111) and Cu(100) model potential, as a function of scaled atom-surface separation (scaled by $n^{-2}$). 

\begin{figure}
\includegraphics[scale=0.45]{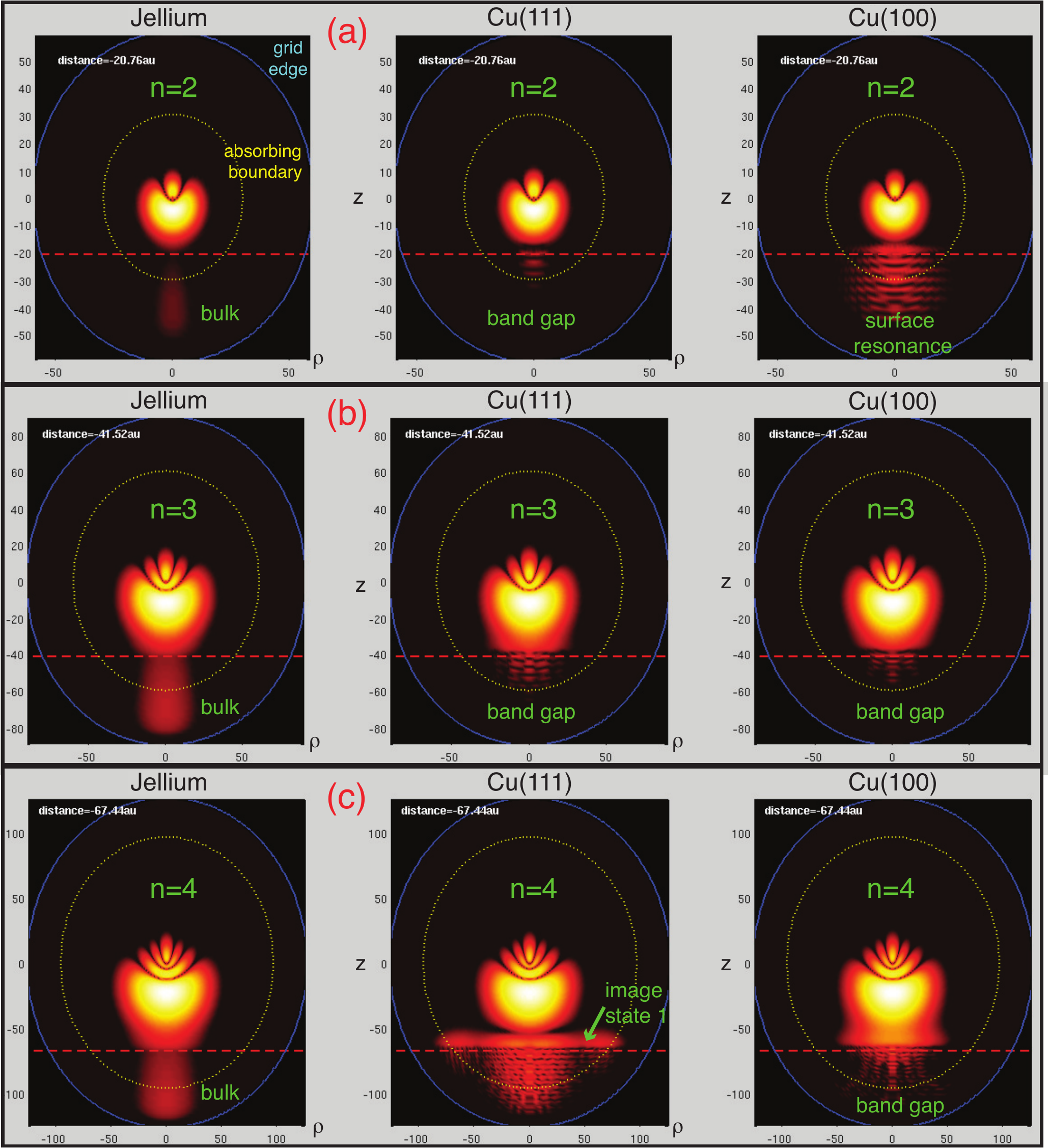}
\caption{Snapshot of the electronic wavefunction, $| \Psi(\mathbf{{r}};t)|^2$, during the wavepacket propagation calculation for the Jellium potential (left panels), Cu(111) potential (middle panels) and Cu(100) potential (right panels) for the (a) $n=2$, (b) $n=3$ and (c) $n=4$ Rydberg states. }
\label{fig:Cun2-4_snapshot}
\end{figure}

For $n=2$, the charge transfer occurs at the greatest distance for Cu(100), then Jellium, and then for Cu(111) at the shortest distance.   Figure~\ref{fig:Cun2-4_snapshot}(a) shows a snapshot of the wavepacket calculation for the three different surfaces; the atom is at a distance of 20.76$a_0$ from the surface.  The amount of electron flux towards the surface clearly follows the order Cu(100) $>$ Jellium $>$ Cu (111).
The $n=2$ Rydberg energy level is nearly degenerate with the surface resonance of the Cu(100) surface at $k_{\parallel} = 0$ ($\bar \Gamma$ point, see Fig.~\ref{fig:dispersion} and Fig.~\ref{fig:Cu_energytable}), and so efficient charge transfer to the surface can occur via the surface resonance with little momentum transfer parallel to the surface -- as illustrated by the directed beam of electron density moving towards the surface.
 The resonant effect is so strong that the charge transfer is even more efficient than for the free-electron Jellium surface. For the Cu(111) surface, the $n=2$  energy is in the energy gap at $k_{\parallel} = 0$, and so charge transfer to the surface state or the 3D bulk states can only occur with large $ k_{\parallel}$ component, which is inefficient since the dominant and preferential charge transfer axis is along $\vec{z}$ where the saddle point in the potential lies. Thus
 little charge transfer has occurred at this point of the trajectory - in accord with the ionization probability profile shown in figure 4.

Figure~\ref{fig:Cun2-8}  shows that for $n=3$, the charge transfer occurs much closer to the metal surface for both the copper surfaces compared with the Jellium. 
This is because the Rydberg energy level is in the energy gap of both Cu(111) and Cu(100) surfaces at the $\bar \Gamma$ point (see Fig.~\ref{fig:dispersion} and Fig.~\ref{fig:Cu_energytable}), and so charge transfer occurs relatively inefficiently with non-negligible ${\bm k}_{{\bm \parallel}}$ components. This behavior  is also apparent in figure 5(b).
For $n=4$, the charge transfer distance is similar for the Cu(111) and Jellium surfaces, while it is much smaller for Cu(100) surface. The snapshot of the wavepacket calculation in Figure~\ref{fig:Cun2-4_snapshot}(c)   for the three different surfaces illustrates that the more significant degree of ionization for the Jellium and Cu(111) surfaces.
The $n=4$  Rydberg energy level is nearly degenerate with the image state of the Cu(111) surface, and so charge transfer can occur via the population of the image state, which can be clearly seen in the horizontal line of electron density, indicated by the arrow in the snapshot in Fig.~\ref{fig:Cun2-4_snapshot}(c). For the Cu(100) surface, the Rydberg energy is within the band-gap and so charge transfer is inefficient.

Above $n=4$ the H atom energies are above the upper edge of the projected band-gap of the Cu(111) surface, and are degenerate with the 3D bulk states (see Fig.~\ref{fig:dispersion} and Fig.~\ref{fig:Cu_energytable}), and so the calculated ionization distance is similar to the Jellium model, as shown in Fig 4 (although resonant transfer to image resonances can still be observed by studying the evolution of the electronic wavefunction, see below). 
\begin{figure}
\includegraphics[scale=0.45]{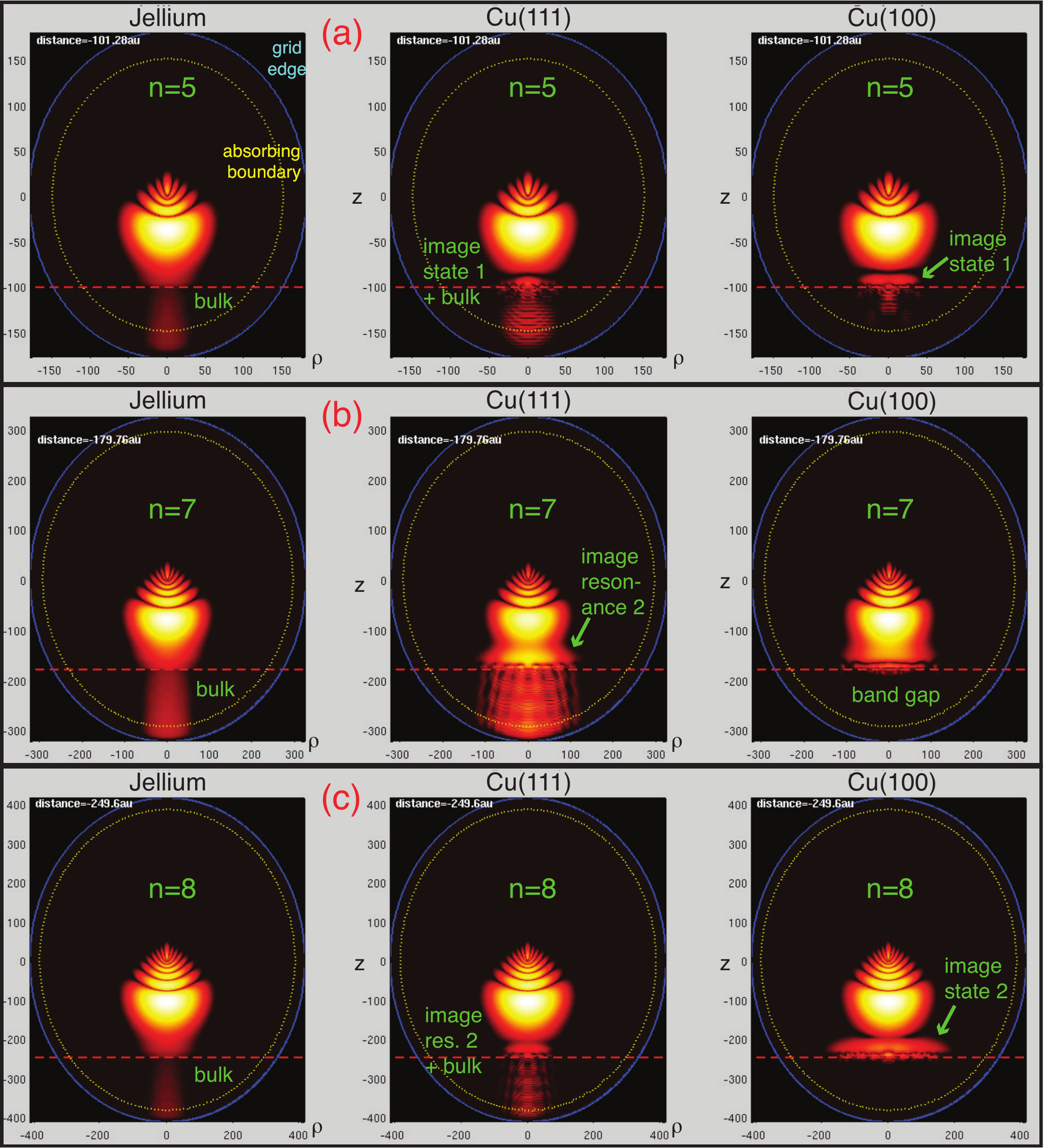}
\caption{Snapshot of the electronic wavefunction, $| \Psi(\mathbf{{r}};t)|^2$, during wavepacket propagation calculation for the Jellium potential (left panels), Cu(111) potential (middle panels) and Cu(100) potential (right panels) for the (a) $n=5$, (b) $n=7$ and (c) $n=8$ Rydberg states. $n=6$ is not shown as the snapshot is qualitatively similar to $n=7$.}
\label{fig:Cun5-8_snapshot}
\end{figure}
For $n=5$, the Rydberg energy is nearly degenerate with the first image state of the Cu(100) surface, and in the projected bandgap, (see Fig.~\ref{fig:dispersion} and Fig.~\ref{fig:Cu_energytable}), and the population of this image state in the charge transfer process can be clearly seen in the snapshot of the wavepacket propagation shown in Fig.~\ref{fig:Cun5-8_snapshot}(a). The resonance charge transfer to the image state results in a similar ionization distance to the Cu(111) and Jellium model surface. This resonance effect to the first image state of Cu(100) is `lost' for $n=6$ and 7, which involves a large ${\bm k}_{{\bm \parallel}}$ transfer, and so the ionization distance becomes progressively smaller than the Cu(111) and Jellium surfaces.
For the Cu(111) surface, ionization of the $n=7$ Rydberg state via the close-in-energy image state resonance (see Fig.~\ref{fig:dispersion} and Fig.~\ref{fig:Cu_energytable}) can be observed in Fig.~\ref{fig:Cun5-8_snapshot}(b). In fact, the effect of the image state resonance is so large that the charge transfer occurs at slightly larger distances than the Jellium surface (Fig.~\ref{fig:Cun2-8}). 

For the Cu(100) surface, the $n=8$ Rydberg energy is close in energy with the second image state embedded in the projected band-gap (see Fig.~\ref{fig:dispersion} and Fig.~\ref{fig:Cu_energytable}). The population transferred to the second image state is illustrated in Fig.~\ref{fig:Cun5-8_snapshot}(c), and the resonance effect is reflected in the ionization probability shown in Fig.~\ref{fig:Cun2-8} (the ionization distance is similar to the Cu(111) and Jellium model surface).

\begin{figure}
\includegraphics{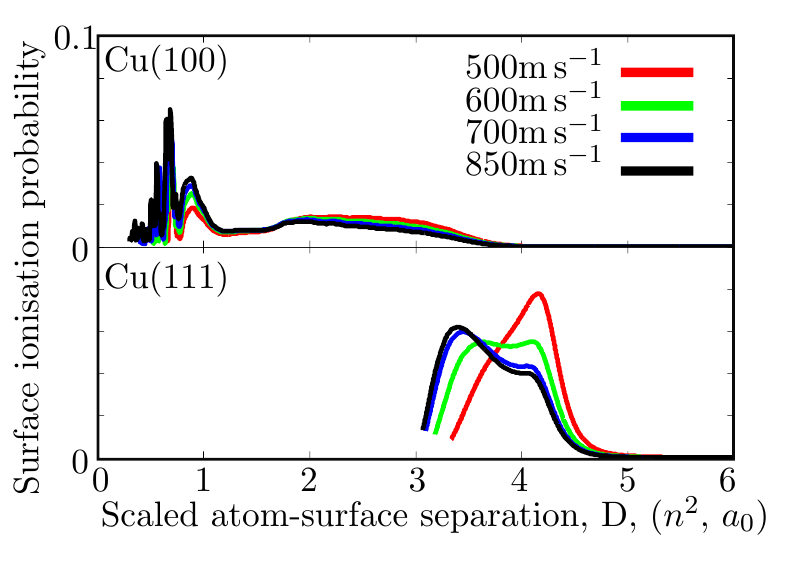}
\caption{The surface ionization probability of the n=4 H atom Rydberg state interacting with (a) Cu(100) and (b) Cu(111) calculated using the constant velocity approximation.}
\label{fig:constant}
\end{figure}
 
 \subsection{Velocity effects}
The wavepacket calculations presented so far in this paper  have used a constant velocity for the incident H atom.  Experimentally we can alter the velocity of the incident H atoms by changing the delay between the pulsed excimer laser that produces H atoms from NH$_3$   and the excitation laser pulses that produce the Rydberg atoms  further down the supersonic beam path \cite{So:PRL}; the velocity can also be varied by altering the seeding gas. Using
unseeded ammonia as the precursor gas we can access a range of perpendicular collisional velocities for the Rydberg H atoms from 500 m/s to 850 m/s while maintaining a usable molecular beam density. Figure \ref{fig:constant} shows repeats of the $n=4$ WPP calculations for the Cu(100) and Cu(111) surface at these typical experimental velocities. In general it would be expected that the higher the velocity, the less time the atom has to ionize in a given incremental range of positions $D\rightarrow D-\delta D$. Therefore the probability distribution for the ionization distance should be shifted to shorter distances at higher velocities.  Second if there are energy level crossings between the Rydberg atom and the surface image states that come into resonance over a specific range of distances, then the crossing would be traversed more quickly at higher velocities and hence the process may be less sensitive to resonance at increasing velocity. 
The results shown in Fig. \ref{fig:constant} 
illustrate these expected trends.
 For the Cu(111) surface in particular, there is a very clear shift of the probability distribution with velocity, with the peak probability shifting from $4.1n^2a_0$ at 500 m/s to around $3.3n^2a_0$ at 850 m/s. 
The shift in the probability distribution is much less pronounced for the Cu(100) surface, as this is a non-resonant case for $n=4$ H atoms, whereas it is resonant for Cu(111). 

In the experiments described in \cite{So:PRL,Cuexpt} there is an additional effect of the velocity impacting on the measurements associated with the probability that the ions produced  are extracted and dectected after ionization.
 For greater collisional velocities the minimum distance of ionization from which the ions  can be extracted for a given field increases, such that the detection region shifts further from the surface. Coupled with the shift in ionization probability to smaller mean atom-surface separations, the integrated ion detection probability will decrease with increasing collisional velocity. 

\begin{figure}
\includegraphics{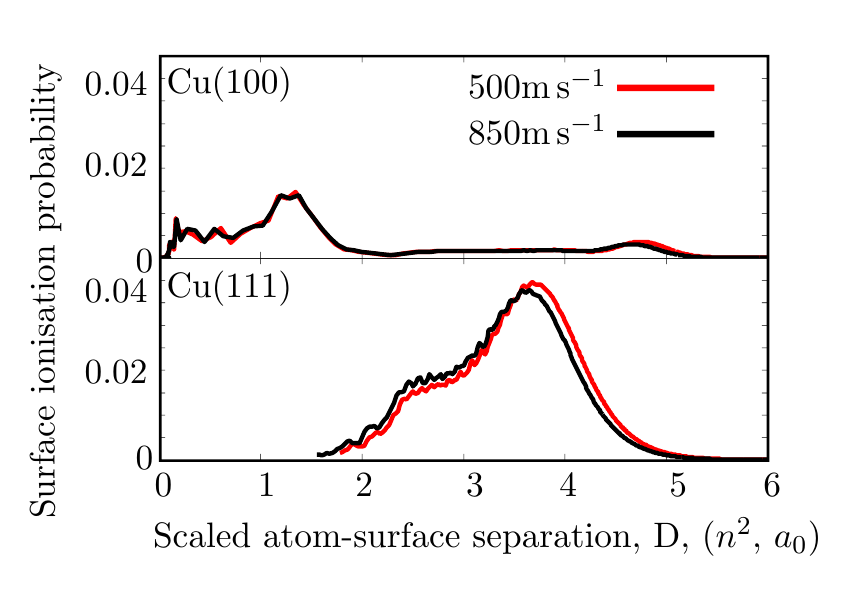}
\caption{The surface ionization probability of the n=4 H atom Rydberg state interacting with (a) Cu(100) and (b) Cu(111) calculated including the mean field term and in the presence of an applied field.}
\label{fig:meanfield}
\end{figure}

In actuality, the velocity of the H atom towards the surface may not be constant, as there is an acceleration due to image-charge attraction. There are effectively three phases of acceleration:
Initially on approaching the surface the  Rydberg atom is accelerated by the van der Waals interaction of the neutral Rydberg atom to the surface, which scales as $D^4$.  In the second phase of the wavepacket model, as the atom starts to ionize it develops an overall partial charge as the electron  flux into the surface is generated, and hence there is a greater acceleration due to the image charge attraction to the surface. Finally the atom becomes fully ionized and there is a classical acceleration of the ion core due to its own image charge.
These effects are neglected in the constant velocity calculations but can be included by the addition of a mean field term \cite{meanfield,So:2009p199}. Here the forces acting upon the ion core are determined, and then the expectation value of the electronic Hamiltonian is calculated on the fly.  
The acceleration will tend to shift the probability of ionization towards lower distances. However, in order to model the experiments a further effect must be included; that of the applied field which is present for extracting the ions.  This will oppose the acceleration of the partially charged atom in the second phase of acceleration. 
In these calculations we use very large (and experimentally impractical) fields, such as would be necessary to detect ions in experiments with $n=4$ Rydberg atoms - in practice experiments are performed with much higher $n$ values and hence much lower, more realistic fields are used.


As shown in figure \ref{fig:meanfield},  inclusion of the mean field term, and the resulting acceleration, causes the expected  reduction in the mean atom-surface separation for ionization.
Additionally the difference between ionization behavior for different incident velocities is reduced, because the acceleration tends to dominate the velocity effects.
For  the on-resonance case, such as the image charge resonance at Cu(111) for $n=4$, the acceleration is less because the ionization occurs further from the surface (and hence image charge effects are less) - thus,  the effects of the initial velocity are still more visible.

The conclusion from the velocity-dependence calculations is that measurements of the velocity dependence of the surface ionization behavior may provide helpful additional indication of resonant behavior.  Recent experimental results from our group indicate that this is indeed the case \cite{Cuexpt}.

\section{Conclusion and future perspective}
The surface ionization dynamics at an electronically structured surface are predicted by these calculations to be very different compared to a free-electron metal surface. Cu(111) and Cu(100) have surface and image states embedded in a projected band-gap along the surface normal, forming a series of 2D bands. The electron motion inside the metal is restricted (based on the energetically available electronic states) in the direction normal to the surface, and is free in the direction parallel to the surface. Since the charge transfer of the Rydberg electron to the metal occurs preferentially along the surface normal (where the saddle point lies), with a minimum component of momentum parallel to the surface, the ionization distance is larger when the Rydberg electron is nearly degenerate in energy with one of the surface and image states. The resonant charge transfer to the surface and image states can be seen directly from the electronic wavefunctions during the wavepacket propagation. 

The charge transfer of Rydberg H atoms with a Cu(100) surface has been studied experimentally in our recent work \cite{Cuexpt}, with principal quantum numbers in the range $n=25-34$, but
 it has not been possible to study the charge transfer process at such low $n$ values as reported in the calculations described here, owing to the short lifetimes of states and large fields required to extract and detect ions. On the other hand, extending the wavepacket calculations beyond $n=10$ would be extremely challenging because of the very large basis sets required. Nevertheless it is possible to make some qualitative predictions for the higher-$n$ experiments.
 
For the Cu(100) surface, the projected band-gap extends to above the vacuum level (upper edge is at +2.98 eV), and the Rydberg series of image states, described by Eq.~\ref{eq:IS_en} with a quantum defect of $ a=0.24$ (and a pre-factor of $1/15.93$ instead of $1/16$) \cite{Klamroth:2001p1713}, will continue throughout this range. Thus, resonance effects are expected to play an important role in the charge transfer dynamics even for the experimental range of principal quantum numbers of $n=20-40$. 

\begin{figure}
\centering
\includegraphics{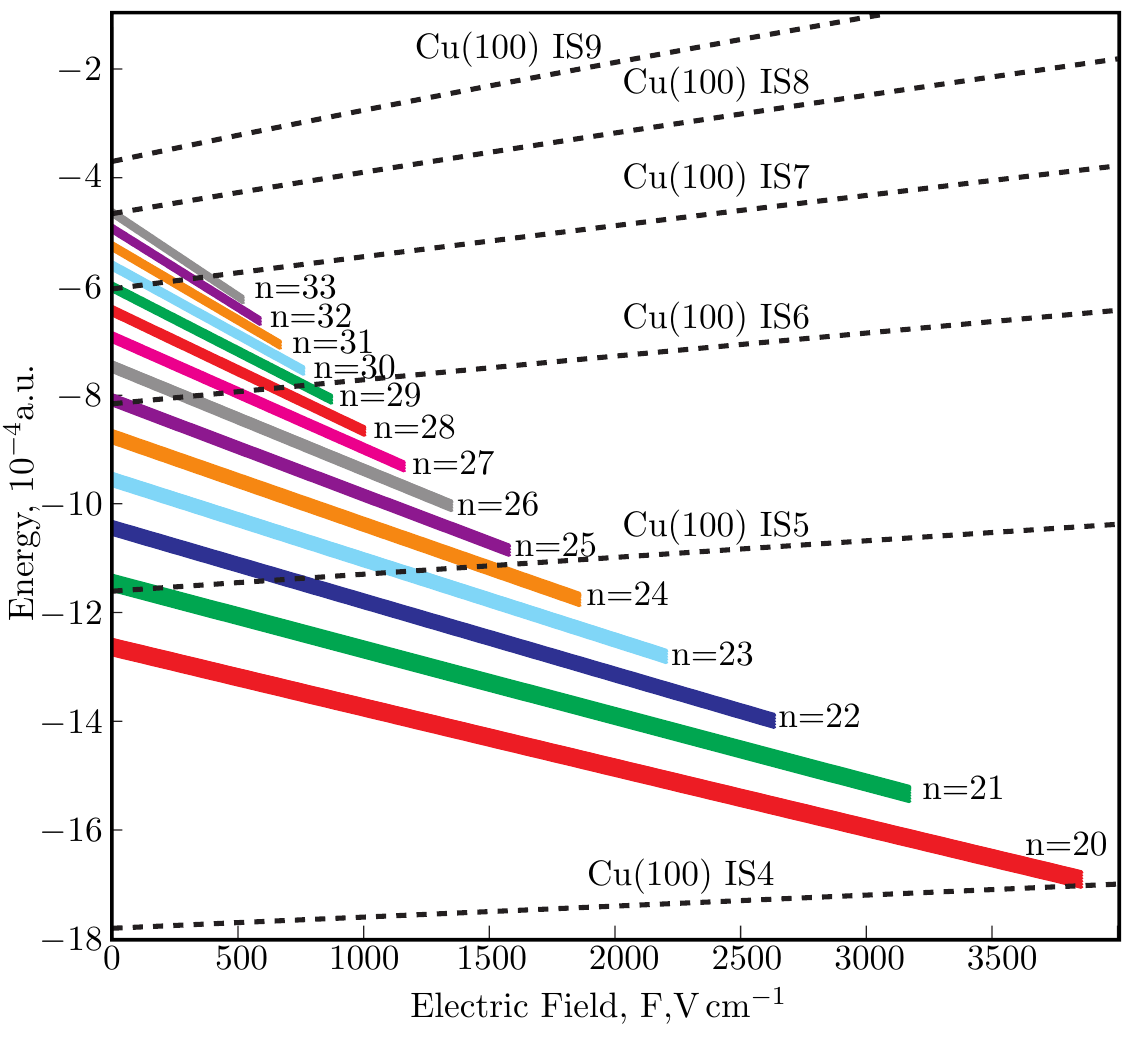}
\caption{ The energies of the H atom Stark states and the energies of the Cu(100) image states as a function of electric field. The broad width of the Rydberg levels shows the approximate shift in energy from $D=\infty$ to $D=3n^2$ a$_0$ calculated using perturbation theory \cite{Borisov:1996p1574}.}
\label{fig:Cu100_red_integrals}
\end{figure}

The image state energies of the Cu(100) surface in zero-field can be calculated from Eq.~\ref{eq:IS_en} with the appropriate quantum defects \cite{Klamroth:2001p1713}. However, it is well known from STM experiments that the energies of image states can be strongly perturbed by external electric fields \cite{BINNIG:1985p1851,Wahl:2003p1850,Crampin:2005p1849,Dougherty:2007p1852}. Following Ref.\ \cite{Wahl:2003p1850}, the Stark shifted energies of the image states are calculated here from the diagonalization of the Chulkov one-electron pseudo-potential given by Eq.~\ref{eq:chulkov}, with the inclusion of a homogeneous electric field beyond the image plane. The calculated energies of the most red shifted $n=20-33$ Stark states of the hydrogen atom, and the $n_{\rm img}=5-9$ image states of the Cu(100), as a function of the electric field are shown in Fig.~\ref{fig:Cu100_red_integrals}(b). The broad width of the Rydberg energies shown in Fig.~\ref{fig:Cu100_red_integrals}(b) represents the approximate shift in energy of the state due to the Rydberg-surface interaction from $D=\infty$ to $D=3n^2$ a$_0$ calculated using first order perturbation theory \cite{Borisov:1996p1574}.
Figure~\ref{fig:Cu100_red_integrals}(b) shows that due the Stark shift of the Rydberg states, most Rydberg levels traverse at least one image-state resonance as the field increases, but that for $n=21,25,29,33$ resonances are expected near zero field with image states IS5 to IS8. $k=0$ Rydberg states, which do not shift in energy with external electric field, may provide the best experimental probe of these high lying image states, as they provide the opportunity for the greatest overlap with the Copper localized image states.

To our knowledge, to date, there has been no direct probe of the $n_{\rm im}\ge4$ image states of Cu(100); in femtosecond two-photon photoemission experiments (where energies of $n_{\rm im}<3$ have been measured), the bandwidth of the laser excited a coherent superposition of high image states ($n_{\rm im}\ge4$) to form wavepackets rather than exciting individual levels \cite{Hofer:1997p1605}. Thus the predictions shown above could lead to some exciting prospects for using Rydberg states as energy specific probes of electronic states at surfaces. 

%

 \bibliography{refs}

\end{document}